\documentclass[
 reprint,
 amsmath,amssymb,citeautoscript,
 superscriptaddress,aip
]{revtex4-2}

\usepackage{dcolumn}
\usepackage{bm}
\usepackage{physics}
\usepackage{comment}
\usepackage{hyperref}
\usepackage{amsfonts}
\usepackage[utf8]{inputenc}
\usepackage{chemformula}
\usepackage{amsmath}
\usepackage{dsfont}
\usepackage{float}
\usepackage{enumitem}
\usepackage[bb=boondox]{mathalfa}
\usepackage[normalem]{ulem}
\newcommand{\kB}{k_\mathrm{B}}

\newcommand{\sbo}{^S\mathrm{Bo}^*}
\newcommand{\tbo}{^T\mathrm{Bo}^*}
\newcommand{\ct}{^S\mathrm{CT}}
\newcommand{\tct}{^T\mathrm{CT}}
\newcommand{\tanth}{^T\mathrm{ANTH}^*}
\newcommand{\wn}{\mathrm{cm}^{-1}}

\begin{document}
\preprint{1}

\title{Solvent Effects on Triplet Yields in BODIPY-Based Photosensitizers}

\author{Leonardo Coello Escalante}
\affiliation{Department of Chemistry, University of California, Berkeley, CA, 94720, USA}

\author{Thomas P. Fay}
\affiliation{Department of Chemistry and Biochemistry, University of California, Los Angeles, CA, 90095, USA \looseness=-1}

\author{David T. Limmer}
\email{dlimmer@berkeley.edu}
\affiliation{Department of Chemistry, University of California, Berkeley, CA, 94720, USA}
\affiliation{Kavli Energy NanoScience Institute, Berkeley, CA, 94720, USA}
\affiliation{\mbox{Materials Sciences Division, Lawrence Berkeley National Laboratory, Berkeley, CA, 94720, USA}}
\affiliation{\mbox{Chemical Sciences Division, Lawrence Berkeley National Laboratory, Berkeley, CA, 94720, USA}}

\date{\today}

\begin{abstract}
We employ molecular dynamics simulations and quantum rate theories to elucidate the complex condensed-phase dynamics underpinning triplet-state formation in organic photosensitizers. Using models informed by first-principles calculations complete with a molecular representation of solvents of different polarities, we elucidate the interplay of the internal and environmental interactions underlying triplet yield. 
We find that triplet yields depend sensitively on the dielectric stabilization of the charge transfer intermediate that facilitates a transition into the triplet manifold. Our results illustrate the importance of molecularly detailed models in understanding the excited-state internal charge-transfer dynamics of photochemically-relevant organic molecules.     
\end{abstract}

\maketitle

\section{Introduction}\label{sec1}
Photoredox catalysts and photosensitizers have emerged as valuable facilitators of unique chemical transformations involving charge or energy transfer. Photocatalysts rely on the formation of long-lived electronic excited states, which most often involves intersystem crossing from a photo-generated singlet state to the triplet manifold in order to exploit the slow, symmetry-forbidden relaxation process back to the ground singlet state. \cite{turro2009principles} The process of intersystem crossing is usually facilitated by spin-orbit coupling,\cite{khudyakov1993spin} and thus the inclusion of heavy atoms in photocatalysts has proven to be a very effective strategy to improve triplet yields. Recently, a promising class of heavy atom-free molecules has emerged, made of donor-acceptor molecular dyads comprised of an electron-deficient boron dipyrromethene (BODIPY) subunit bonded at the \textit{meso} position in a non-coplanar geometry to an electron-rich, usually aromatic subunit \cite{bassan2021design, klfout2017bodipys}. These molecules have been observed to achieve efficient generation of triplet states through a singlet charge-transfer intermediate, in a process known as spin-orbit charge-transfer intersystem crossing (SOCT-ISC)\cite{dance2008intersystem}. The excited-state dynamics of these new systems are not well understood, hampering their design. Here, we use molecular models and quantum rate theories to elucidate the myriad factors that govern the triplet yields of BODIPY-based photosensitizers. 

The efficiency of a triplet photosensitizer is measured by its triplet yield and determined by a delicate interplay of several thermodynamic and kinetic factors, each of which depends sensitively on the surrounding condensed phase environment. While the inherent electronic structure of the molecule establishes the feasibility of triplet-state formation, specific molecular interactions and thermal fluctuations ultimately set the favorable free-energy differences that facilitate transitions and determine the timescales associated with activated excited-state dynamics. An example is the contrasting trends in triplet yield as a function of solvent polarity exhibited by BODIPY-based molecular dyads with different donor subunits. \cite{uddin2023twist} In particular, a BODIPY-anthracene derivative (BoANTH) was reported to have a high singlet oxygen yield, used as a lower-bound to the triplet yield, of 0.76 when dissolved in acetonitrile (ACN), but considerably poorer yield of 0.38 when dissolved in toluene (TOL). Conversely, a BODIPY-N-methyl phenothiazine complex (BoPTH) showed the opposite trend, with a vanishingly small singlet oxygen yield, $<$0.01, in ACN, but a surprisingly large yield of 0.73 in TOL. While simple dielectric continuum arguments offer tentative explanations of this effect, a definitive account requires detailed molecular insights. Thus, a reliable theoretical description of the kinetic processes involved in photocatalytic activation, which accounts for the specific effects of the solvent-dyad interactions, is a desirable tool to interpret experimental results and inform design strategies for novel catalysts.

\begin{figure}
    \centering
    \begin{tabular}{c}
         \includegraphics[width=\linewidth]{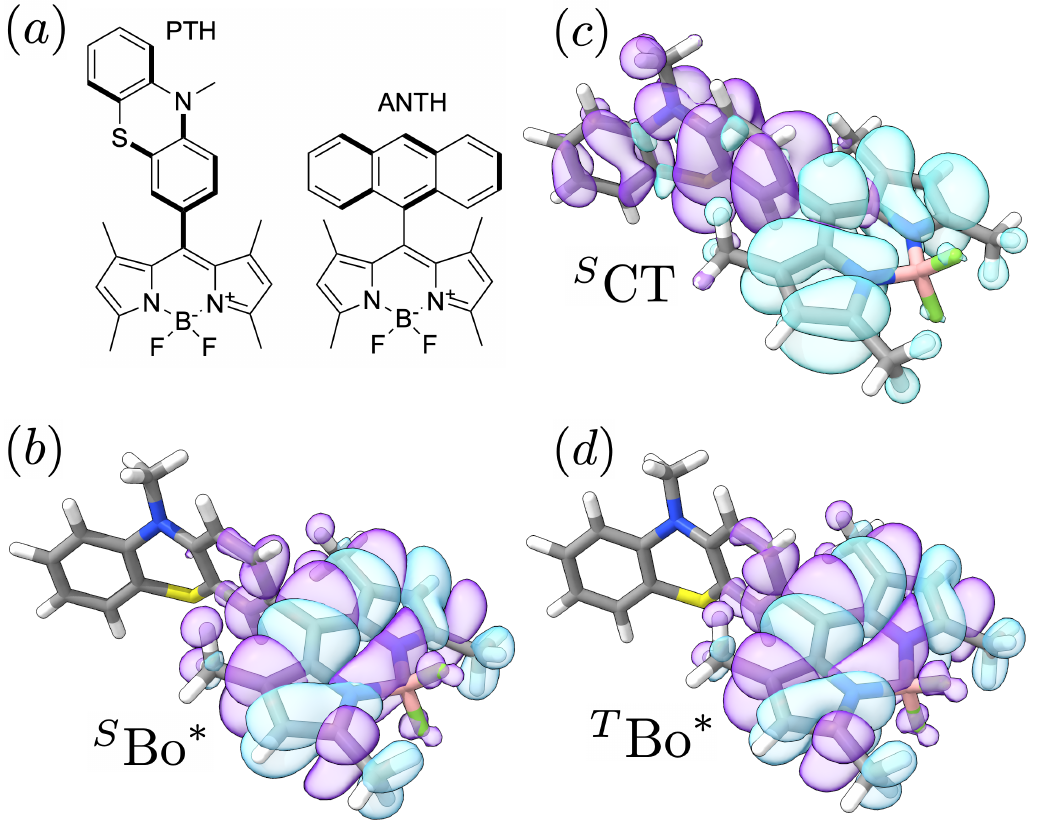} \\
    \end{tabular}
    \caption{(a) Structures of the sensitizers considered in this work, BoPTH (left) and BoANTH (right). (b)-(d) Difference density plots of the exicted states in BoPTH, showing the character of the electronic excitation. }
    \label{fig:elec_struct}
\end{figure}

To clarify the factors that determine triplet yield in BODIPY dyads, we adopt an approach based on a spin-boson mapping to the Fermi's Golden Rule rate for transitions between electronic states. This mapping approach has been shown to yield good estimates of quantum rates through the construction of realistic spectral densities from atomistic simulations.\cite{warshel1986simulation, hwang1997relationship, bader1990role, blumberger2015recent, lawrence2020confirming} 
 We build on recent developments that allow for the efficient construction of bespoke excited-state molecular dynamics forcefields with state energies and frequencies derived from electronic structure calculations to realize these calculations.\cite{fay2024unraveling} Our approach produces results that are qualitatively consistent with experimental observations, and are able to capture relevant mechanistic details associated with solvation that can only be attained by a fully atomistic and large-scale simulation of the condensed phase system. 
 We find that the low triplet yield of BoANTH in TOL is largely a consequence of the charge-separation process becoming thermodynamically unfavorable compared to charge separation in ACN, while in BoPTH this process remains thermodynamically feasible in both solvents. For both dyads, the triplet yields depend sensitively on the excited state lifetime, which is set by the radiative decay rate from photoexcited state and the nonradiative charge recombination from the charge transfer state, which lies deep in the Marcus inverted regime.

This paper is organized as follows. In Section \ref{sec2}, we present details related to electronic structure calculations, as well as diabatic state and forcefield construction. In Section \ref{sec3}, we present the theoretical background underpinning the calculation of rates and spectra from a spin-boson mapping, validate the approach via the calculation of absorption lineshapes, and discuss the insights into solvent effects that can be extracted from spectral densities. In Section \ref{sec4}, we present the calculated rates governing the excited-state dynamics, and we calculate and discuss the associated triplet yields, before providing concluding remarks in Section \ref{sec5}.

\section{Electronic structure}\label{sec2}

Our molecular description is built on \textit{ab initio} calculations. In the first part of this section, we discuss the mixture of time-dependent density functional theory (TDDFT) and higher-level wavefunction methods that we used to identify energetically-relevant excited states, and calculate electronic excitation energies,  equilibrium geometries and frequencies. Subsequently, we elaborate on how we used this information to construct diabatic representations of the excited states, yielding couplings for the calculation of Fermi's Golden rule rates, and to build molecular forcefields through a least-squares optimization procedure. Throughout we consider both BoANTH and BoPTH molecular dyads, whose structure is illustrated in Fig.~\ref{fig:elec_struct}a).  

\subsection{Excited state manifold}
\begin{figure}[t]
    \centering
         \includegraphics[width=\linewidth]{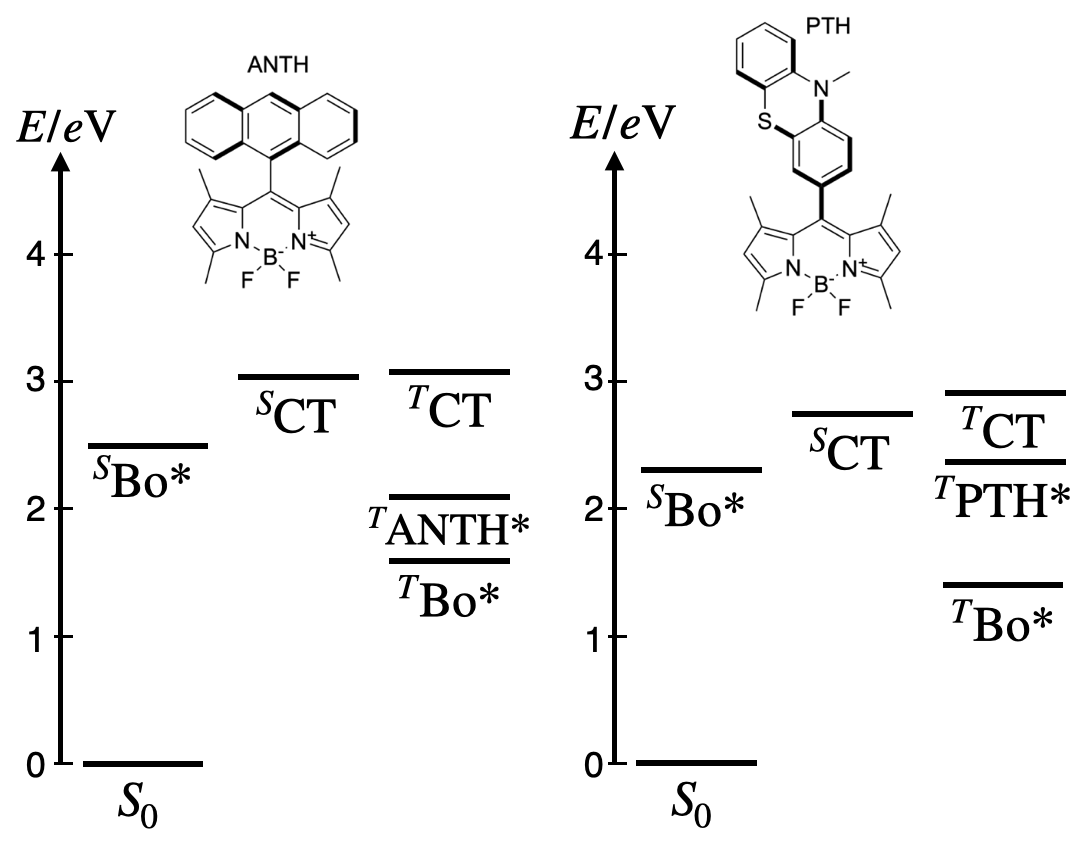} \\
    \caption{Gas phase energies (in $e$V) of the different excited states of both molecular dyads at their respective minimum energy geometries (from implicit solvent(ACN) TDA-TDDFT optimizations), computed at the DLPNO-STEOM-CCSD level of theory}
    \label{fig:rate_data}
\end{figure}

All electronic structure calculations were performed using the Orca 6.0 software package \cite{neese2012orca, neese2020orca, neese2025software}.
In BoPTH, only four electronic states were found to be mechanistically relevant, the ground singlet state $S_0$, the first excited singlet state $\sbo$, the associated singlet charge transfer state $\ct$, and the lowest energy triplet state $\tbo$.
In related molecules including BoANTH, higher-lying triplet states such as a donor-localized triplet and a charge-transfer triplet state depicted in Fig. \ref{fig:rate_data} have been demonstrated to play an important role in triplet state formation. Experimentally, the $\tct$ state has also been observed in dyads related to those that we study here\cite{dong2019spin}. Our calculations show, however, that for BoPTH the increased stability of the $^S\mathrm{CT}$ state renders ISC into these higher triplets thermodynamically unfavorable. This, alongside large reorganization energies, makes their role in the photophysics negligible. A more detailed discussion of the calculations justifying this decision can be found in Appendix A.
Figures \ref{fig:elec_struct}(b)-(d) show difference density plots, the difference between ground and excited state electronic densities, illustrating that the first singlet excited state corresponds to an excitation predominantly localized in the BODIPY fragment, whereas the second singlet state is clearly the result of a charge-transfer excitation. This is further confirmed by noting that this excitation is dominated by a transition where the donor and acceptor molecular orbitals are localized on the PTH and BODIPY fragments, respectively. Further, the gas-phase dipole moment of this state is 23.2 D, over three times larger than the corresponding ground state dipole moment (7.2 D), indicating a significant degree of charge-separation. State geometries and Hessians were calculated with TDA-TDDFT, using the rev-M11 range-separated hybrid functional, which has been shown to perform well in problems with charge-transfer excitations \cite{verma2019revised, verma2020status, lehtola2018recent}, and the def2-TZVP basis set.\cite{weigend2005balanced,weigend2006accurate} The electronic excitation energies and transition dipole moments were calculated from DLPNO-STEOM-CCSD and def2-SVP basis set. The geometry optimizations and frequencies were calculated in implicit solvent (ACN) while the electronic energies were calculated without implicit solvent.

For BoANTH, a total of six electronic states were identified as energetically relevant in a previous study \cite{fay2024unraveling}. Here, we will use the quantum chemistry data for BoANTH reported in that work. The energetically relevant singlet states are the ground electronic state $S_0$, the first singlet excited state $\sbo$, corresponding to an excitation localized in the BODIPY fragment, and the singlet charge transfer state $\ct$, in which charge is moved from the anthracene donor onto the BODIPY acceptor. The relevant triplet states, as alluded to previously, are the lowest energy triplet that features a localized excitation in the BODIPY fragment, followed by a triplet charge transfer state $\tct$, and a triplet state with the excitation localized in the anthracene unit, $\tanth$. In a previous study\cite{fay2024unraveling}, it was found that the transition between singlet and triplet charge transfer states contributed somewhat to triplet state formation, but to a much more limited extent owing to the fact that the transition violates El-Sayed's rule. Ground and excited state geometries were optimized and their corresponding Hessians were calculated with TDA-TDDFT\cite{runge1984density, hirata1999time}, using the $\omega$B97X-D3 range-separated hybrid functional \cite{lin2013long} and the def2-SVP basis set \cite{weigend2005balanced}. More accurate estimates of the electronic excitation energies and transition dipole moments were obtained from higher level wave-function based calculations, at the DLPNO-STEOM-CCSD/def2-TZVP(-f) level of theory \cite{dutta2016towards, dutta2018exploring}. The energies for the relevant states in both molecules are shown in Fig.~\ref{fig:rate_data}. We have verified that the two functionals employed in this work produce consistent values of normal mode frequencies and commensurate equilibrium geometries, leading to similarly consistent vibronic couplings and spectral densities.

\subsection{Diabatic state potentials}
While electronic structure calculations give access to locally-adiabatic states, transitions between excited states are more aptly described within a diabatic representation. Diabatic states were constructed and inter-state couplings were calculated from the adiabatic state data following the generalized Mulliken-Hush procedure \cite{cave1996generalization}, which relies on the diagonalization of the dipole moment operator projected along the charge-transfer axis. The spin-orbit couplings mediating intersystem crossing to the triplet manifold were calculated using the mean-field approach to spin-orbit coupling implemented in Orca \cite{neese2005efficient, de2019predicting}. This procedure was carried out separately in the equilibrium geometries of all relevant electronic states of the molecular sensitizers. The couplings agree reasonably well at different geometries, justifying the use of the Condon approximation, and the final values of coupling were obtained as an average 
\begin{equation}
    |V_{AB}|^2 = \frac{1}{2}\left(\left|V^{(A)}_{AB}\right|^2+\left|V^{(B)}_{AB}\right|^2\right)
\end{equation} 
where $V^{(A)}_{AB}$ denotes the coupling evaluated at the equilibrium geometry of A. 
We find that rev-M11 and $\omega$-B97X-D3 provide consistent spin orbit couplings in a given geometry, with differences on the order of $10^{-2}$ cm$^{-1}$. An explicit comparison of the couplings evaluated at different equilibrium geometries can be found in Appendix B.

The couplings, shown in Table \ref{tab:all_data}, exhibit similar trends across the two molecules, with the charge recombination to $S_0$ having by far the largest, followed by the charge separation transition between $\sbo$ and $\ct$. The couplings between $\ct$ and $S_0$ are so strong that higher order effects in perturbation theory beyond Fermi's Golden Rule could be relevant. To stay within the purview of nonadiabatic rate theory, we opt to apply corrections to the Golden Rule rate constant to approximately account for these higher order processes. In particular, we apply the ``Optimal Golden Rule" correction (OGR), which relies on calculating rates in a rotated basis that minimizes the diabatic coupling \cite{fay2024extending}. This is discussed in greater detail in Section \ref{sec4}. Spin-orbit couplings, even for SOCT-ISC transitions obeying El-Sayed's rule, are significantly smaller, on the order of 0.7 $\wn$. 

While most of the state energies are minimally modified by the basis transformation, this is not the case for the $\ct$ state of BoPTH, which is significantly stabilized upon diabatization in its equilibrium geometry. In particular, we find that this state is lowered by nearly 0.5 $e$V with respect to its adiabatic energy, putting it slightly lower in energy compared to the corresponding diabatic $\sbo$ state. This has clear implications for the thermodynamics of charge separation, discussed in the following sections.

We constructed bespoke forcefields for the electronic states of the molecular dyads as well as the solvents, through the use of a recently developed Hessian fitting protocol\cite{fay2024unraveling} based on the OPLS-AA forcefield ansatz \cite{jorgensen1996development, dodda2017ligpargen}. The potential consists of harmonic local stretching and bending modes linearly coupled to first neighbors, as well as proper and improper dihedrals, Lennard-Jones and Coulomb interactions
\begin{multline}
    U  = U_\mathrm{bond} + U_\mathrm{angle} + U_\mathrm{bond-bond} + U_\mathrm{bond-angle} + \\
    U_\mathrm{torsion} + U_\mathrm{LJ} + U_\mathrm{Coul}
\end{multline}
whose forms are given in Appendix C. The procedure to determine the parameters in this forcefield consists of a partial Hessian fitting, through a quadratic-loss optimization, of the normal mode quantum-mechanical Hessian projected into local stretching, bending and torsional degrees of freedom. The loss function has the form
\begin{equation}
    \mathcal{L}_\mathrm{Hess} = \sum_{A\geq B} \|\mathbb{H}_{AB}^{(\mathrm{QM})} - \mathbb{H}_{AB}^{(\mathrm{MD})}\|^2
\end{equation}
where $\mathbb{H}_{AB} = \nabla_A\nabla_B^\mathrm{T} U$ is the $3\times 3$ partial Hessian block of atoms $A$ and $B$ and superscripts refer to the forcefield(MD) or ab initio Hessian (QM). During the first stage of fitting, bond lengths and angles are allowed to change freely, but they were subsequently refined by minimizing another loss function for the difference between the equilibrium geometries and the quantum mechanical reference geometries. The fittings were achieved with typical root-mean-squared errors of $\sim 30 - 50 \ \wn$ in the normal mode frequencies, and of $\sim 0.1 - 0.2 \ \AA$ in the geometry. Optimizing the frequencies of the forcefield to match those of the electronic structure calculations helps to ensure that the resultant estimates of inner-sphere reorganization energies will be accurately modeled. 

The atomic partial charges were assigned as an average of the gas phase and CPCM(ACN) charges obtained from quantum mechanical calculations \cite{mulliken1955electronic}. 
We employ this averaging procedure as it has been observed previously to better approximate solvation energies for non-polarizable forcefields.\cite{muddana2014sampl4, jorge2023optimal} 
CHELPG charges were used in BoANTH, and Mulliken charges were used in BoPTH \cite{breneman1990determining}. From theoretical arguments and numerical simulations, it is well-known that electronic polarizability usually has the effect of attenuating reorganization energies \cite{song1993quantum, marchi1993diabatic, blumberger2008reorganization}. As a means to include solvent polarizability, we worked with polarizable models of ACN and toluene based on the Drude oscillator model. Polarizability effects were not included in the forcefields of the molecular dyads. 

The electronic structure of the charge-transfer state in BoPTH proved to be challenging. In the gas phase, the equilibrium configuration of $\ct$ was found to involve a significant conformational change, where the phenothiazine subunit went from a puckered to a planar geometry. However, geometry optimizations targeting this state in implicit solvent systematically failed to converge to a state with charge-transfer character. Subsequent potential energy scans revealed that in a higher dielectric environment, this state exhibited conical intersections with the lower-energy singlet state along the puckering coordinate, likely explaining the inability to optimize it in implicit solvent. On the premise that the $\ct$ state has been experimentally observed in solution, we opted to construct an effective model for the $\ct$ state with the appropriate charge distribution, by using the (puckered) geometry optimized in CPCM while still using an average of the gas-phase and CPCM charges for the forcefield. This choice was also made in consideration of the limitations of the spin-boson mapping employed in this work, which is known to break down for transitions involving dramatic conformational changes \cite{blumberger2006diabatic, krapf2012road}. The diabatization was still done with the transition dipoles and energies evaluated at the planar gas phase geometry of the true charge transfer state.  

\section{Nonadiabatic transitions rates}\label{sec3}
Having established the relevant electronic states and their associated potential energy surfaces, we now consider a means by which we can approximate their quantum dynamics. We will assume that the nuclear environment responds linearly to perturbations from the electronic degrees of freedom. This is a valid assumption for intramolecular changes accompanying a transition for rigid molecules and also for describing the response of collective degrees of freedom, like solvent polarization in response to a charge transfer. We further assume that we can consider transitions between individual pairs of states at a time and evaluate their rates with perturbation theory, or corrections thereof. 

Within these assumptions, for a pair of electronic states, $A$ and $B$, we can define a two-state Hamiltonian coupled to a bath of nuclear coordinates ($\mathbf{\hat{Q}}$) that include both intramolecular vibrational modes and outer-sphere solvent degrees of freedom with conjugate momentum $\mathbf{\hat{P}}$. In a diabatic representation, the Hamiltonian $\hat{H}$ is
\begin{equation}
    \label{Hamiltonian_general}
    \hat{H} =\begin{bmatrix}  \hat{H}_A(\hat{\textbf{P}},\hat{\textbf{Q}}) & \hat{V}\\ \hat{V} & \hat{H}_B(\hat{\textbf{P}},\hat{\textbf{Q}})
    \end{bmatrix}
\end{equation}
where the electronic states are connected by a coupling $\hat{V}$ that we take to be a constant, under the traditional Condon approximation.\cite{condon1928nuclear} The electronic state-specific nuclear Hamiltonians, $\bra{\alpha}\hat{H}\ket{\alpha}=\hat{H}_\alpha(\hat{\textbf{P}},\hat{\textbf{Q}})$, for $\alpha=\{A,B\}$ can be generically expressed as
\begin{equation}
    \label{diab_Ham}
    \hat{H}_\alpha(\hat{\textbf{P}},\hat{\textbf{Q}}) = \hat{K}(\mathbf{\hat{P}}) + \hat{U}_\alpha(\mathbf{\hat{Q}})
\end{equation}
where $\hat{K}(\mathbf{\hat{P}})$ is the kinetic energy of the nuclei, and $\hat{U}_\alpha(\mathbf{\hat{Q}})$ denotes the full-dimensional potential energy surface of electronic state $\alpha$. 

The flux correlation function between the states of interest, $c_{A,B}(t)$, contains all the dynamical information to describe thermal transition rates and optical spectra. For sufficiently small diabatic coupling, it is appropriate to take the non-adiabatic Fermi's Golden Rule limit, leading to a flux correlation function quadratic in $V$,\cite{limmer2024statistical}
\begin{equation}
    \label{flux_tcf}
    c_{A,B}(t) = \frac{|V|^2}{\hbar^2} \left  \langle e^{-i\hat{H}_A t/\hbar} e^{i\hat{H}_B t/\hbar} \right \rangle_A  
\end{equation}
where $\langle\dots\rangle_A$ denotes a thermal ensemble average over nuclear degrees of freedom evaluated with $\hat{H}_A$ 
\begin{equation}
        \langle\dots\rangle_A = \frac{\Tr_N\left[e^{-\beta\hat{H}_A}(\dots)\right]}{\Tr_N\left[e^{-\beta\hat{H}_A}\right]}
\end{equation}
where $\beta = 1/\kB T$ is the inverse of Boltzmann's constant times the temperature. 

The product of propagators in Eq. \ref{flux_tcf} can be expressed as a single time-ordered exponential, allowing $c_{A,B}(t)$ to be expressed in terms of a generalized cumulant expansion of the energy gap operator, $\hat{\Delta}\equiv \hat{H}_A - \hat{H}_B$. Under the assumption that the bath responds linearly, 
\begin{multline}
    \label{cum_exp}
    \left \langle e^{-i\hat{H}_A t/\hbar} e^{i\hat{H}_B t/\hbar} \right  \rangle_A \approx  \\ 
    \exp\left[-\frac{it}{\hbar}\langle\hat{\Delta}\rangle_A
    -\frac{1}{\hbar^2}\int_0^t(t-\tau)\langle \delta \hat{\Delta}(0) \delta\hat{\Delta}(\tau)\rangle_A d\tau\right] 
\end{multline}
we can truncate this series at second order, including only the first and second cumulants of the time-integrated energy gap\cite{kubo1955application}.
This truncation allows us to express the flux correlation function purely in terms of the spectral density of the system, $\mathcal{J}_{AB}(\omega)$, which is connected to the energy gap correlation function through
\begin{equation}
    \label{spec_dens_tcf}
    \langle\delta\hat{\Delta}(0)\delta\hat{\Delta}(t)\rangle_A = \frac{\hbar}{\pi}\int_{-\infty}^{\infty}  \frac{e^{-i\omega t}}{1-e^{-\beta\hbar\omega}}\mathcal{J}_{AB}(\omega) d\omega
\end{equation}
where the Bose occupation factor in the denominator of the integrand originates from the fact that $\mathcal{J}_{AB}(\omega)$ is defined with respect to the Kubo transform of the energy gap correlation function \cite{kubo1957statistical}. Inserting Eq. \ref{spec_dens_tcf} into \ref{cum_exp} leads to a compact expression of the flux correlation function \cite{nitzan2024chemical} 
\begin{equation}
    \label{SB_flux_tcf}
    c_{A,B}(t) = \frac{|V|^2}{\hbar^2} e^{\left [i\Delta F_{A, B}t - \chi'(t) -i\chi''(t)\right ]/\hbar}
\end{equation}
where the free energy difference between states $A$ and $B$ is given by
\begin{equation}
    \label{deltaF_LR}
    \Delta F_{A,B} = \langle\hat{\Delta}\rangle_A + \lambda 
\end{equation}
with reorganization energy
\begin{equation}
    \lambda_{AB} = \frac{1}{\pi}\int_0^\infty \frac{\mathcal{J}_{AB}(\omega)}{\omega}d\omega 
\end{equation}
and 
\begin{equation}
    \chi'(t) = \frac{1}{\pi} \int_{0}^{\infty} \frac{\mathcal{J}_{AB}(\omega)}{\omega^2}\frac{1-\cos(\omega t)}{\tanh\left(\beta\hbar\omega/2\right)}d\omega
\end{equation}
\begin{equation}
    \chi''(t) =\frac{1}{\pi} \int_{0}^{\infty} \frac{\mathcal{J}_{AB}(\omega)}{\omega^2}\sin\left(\omega t\right) d\omega 
\end{equation}
are the real and imaginary components of the gap correlation function. 

In order to evaluate $c_{A,B}(t)$ for pairs of states, we need to compute the time correlation function of the energy gap fluctuations. To do this, we assume that we can replace the quantum mechanical   $\langle\delta\hat{\Delta}(0)\delta\hat{\Delta}(t)\rangle_A$ with its classical limit, $\langle\delta\Delta(0)\delta\Delta(t)\rangle_A$. Since these correlation functions are Kubo transformed, for a linearly responding bath this is expected to be a good approximation. Employing this approximation means that the spectral density of the system can be sampled from classical molecular dynamic simulations. It is equivalent to mapping the atomistic system to a spin-boson model by sampling energy gap correlations with classical molecular dynamics of a realistic representation of the true, anharmonic system, and then assuming that the spectral density associated to this classical correlation function is the same as that of a fictitious spin-boson system, for which the second-order cumulant expression is exact. 

With  $c_{A,B}(t)$ extractable from molecular simulation, the nonadiabatic rates of transitions, $k^{(\mathrm{NR})}_{A,B}$, between states $A$ and $B$ can be computed directly as a time integral,
\begin{equation}
    k^{(\mathrm{NR})}_{A,B}(\Delta F_{A,B}) =  \int_{-\infty}^{\infty}  c_{A,B}(t) dt
    \label{nonrad_rate}
\end{equation}
where it will be useful to compute these rates as a function of the driving force. Further, the linear absorption spectrum is given by the Fourier transform of the flux correlation function 
\begin{equation}
    I(\omega) = \frac{1}{2\pi}\int_{-\infty}^{\infty} e^{-i\omega t} c_{A,B}(t) dt
\end{equation}
if $A$ is taken to be the ground state, and $B$ the first singlet excited state \cite{zuehlsdorff2019optical,wiethorn2023beyond}. This approach gives the correct Marcus result in the classical limit. Provided that the forcefields accurately represent the system, and anharmonic bath modes are not dominant contributors to activated processes, it has been shown to provide reasonable estimates of non-adiabatic rates, that approximately captures nuclear quantum effects \cite{warshel1986simulation, hwang1997relationship, bader1990role, blumberger2015recent, lawrence2020confirming}. 

\begin{figure}[t]
    \centering
    \includegraphics[width=0.9\linewidth]{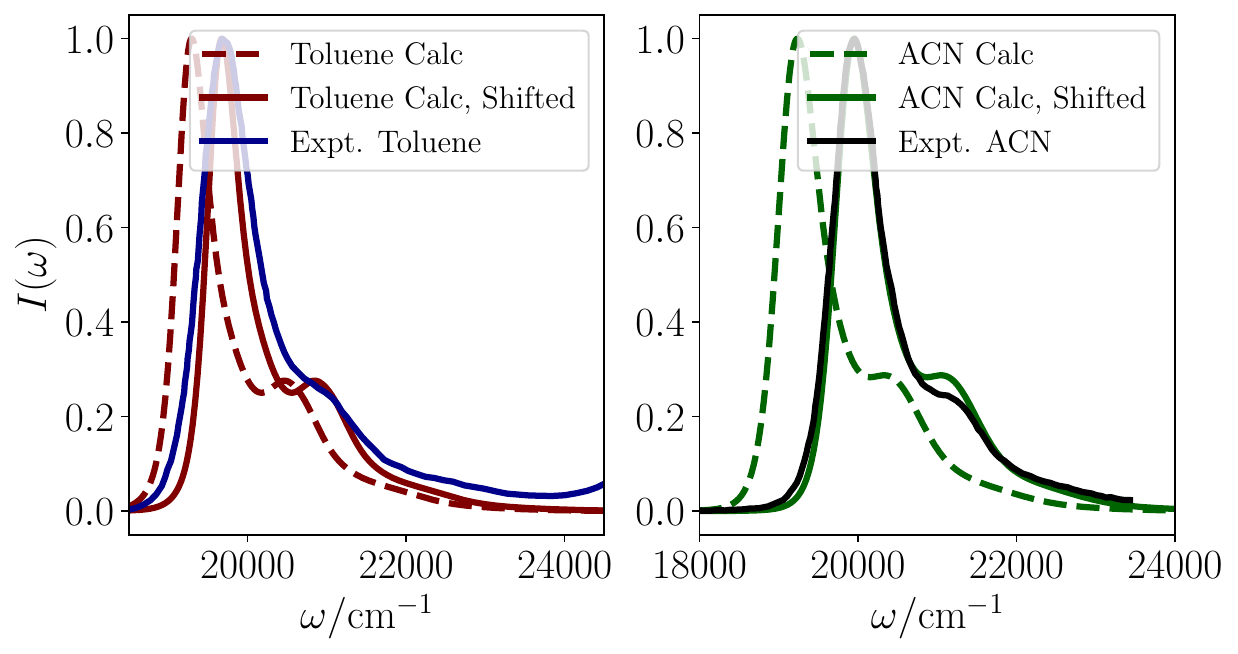}
    \caption{Simulated and experimental absorption lineshapes of BoANTH in toluene and ACN}
    \label{fig:BoANTH lineshape}
\end{figure}
\begin{figure}
    \centering
    \includegraphics[width=0.9\linewidth]{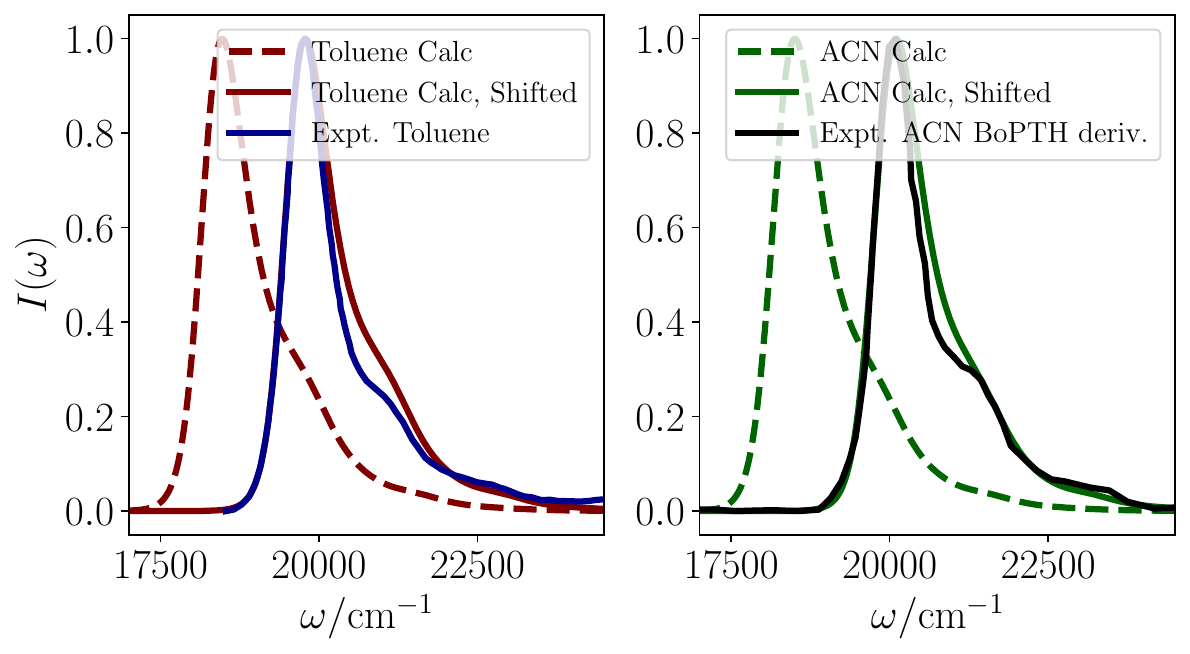}
    \caption{Simulated and experimental absorption lineshapes of BoPTH in toluene and ACN}
    \label{fig:BoPTH lineshape}
\end{figure}

\subsection{Spectral Distributions and  Absorption Spectra}
The bespoke state-specific forcefields were calibrated by calculating the condensed-phase absorption lineshape of the two molecules in ACN and TOL, displayed in Figs. \ref{fig:BoANTH lineshape} and \ref{fig:BoPTH lineshape}. In lieu of experimental absorption data for BoPTH in ACN, the simulated lineshape for this system is shown alongside the spectrum of a derivative of BoPTH, in which the methyl group in N-methyl phenothiazine is replaced by a butyl group \cite{chen2017triplet}. This is expected to be a reasonable comparison because the absorption spectrum in these molecular dyads is dominated by the BODIPY fragment, owing to the fact that the electronic excitation of the first singlet excited state is localized in this region of the molecule. All of the calculated lineshapes successfully capture the vibronic structure of the spectra. Most notably, there is a small shoulder in both molecules roughly 1200 $\wn$ above the main absorption peak that is also present in the simulated spectra, meaning that the frequencies obtained from the Hessian fitting procedure are sufficiently accurate to reproduce the vibronic coupling effects that are observed experimentally. The reorganization energies in BoPTH overestimated the broadening, so they were scaled down by $40\%$ in both solvents.            
These spectra also give an opportunity to evaluate the accuracy of the calculated electronic gaps, by comparing the location of the main absorption peak to experiment. Without shifting the spectra, the gap between $S_0$ and $\sbo$ is systematically underestimated by up to $\sim$0.1 $e$V, likely indicating errors in the electronic structure calculations, given that EOM-CCSD is known to give average errors for charge-transfer excitations of $\sim$ 0.3 $e$V. Previous studies\cite{fay2024unraveling} have found that increasing the basis set size from def2-SVP to def2-TZVP(-f) results in a blue shift of the $S_0 - \sbo$ gap of $\sim$0.1 $e$V, but the change in the energy difference between the excited states is much smaller. In light of this observation, we have shifted the energy of all the excited states by the difference between the main peak in the simulated and experimental spectra. For BoANTH, where we have access to the spectra in both solvents, the smallest shift of the two was chosen, resulting in a shift of 237.7 $\wn$ for BoANTH, and 1312.5 $\wn$ for BoPTH.

Having verified that the forcefields yield acceptable agreement with experimental spectra, we now present the spectral densities for the other relevant transitions. To compare different solvents, it is useful to define a spectral distribution function, $\rho_{AB}(\omega)$ as
\begin{equation}
    \label{spec_distr}
    \rho_{AB}(\omega)= \mathcal{J}_{AB}(\omega)/\pi\lambda_{AB}\omega
\end{equation}
which normalizes the spectral density by the reorganization energy. 
The spectral densities were constructed using the classical limit of Eq. \ref{spec_dens_tcf}, from a corresponding energy gap correlation function sampled in molecular simulation from 20 independent NVT ($T = $ 298 K) trajectories of 30 ps in length, where the energy gap was evaluated every 0.5 fs after 1 ns of equilibration. In all cases, the system consisted of the molecular dyad embedded in a periodic box with 1000 solvent molecules, that had been previously equilibrated in the NPT ensemble for 1 ns. Trajectories were evolved through the velocity Verlet algorithm with a Langevin thermostat with friction coefficient of 2.0 ps$^{-1}$, and long-range electrostatics were calculated with the PPPM method. All simulations were carried out using the OpenMM 7 software package \cite{eastman2017openmm}. These trajectories were obtained using non-polarizable forcefields for the solvent as well as the photosensitizers. To correct for polarizability effects, we weighted the contributions of the inner and outer sphere components of $\rho(\omega)$ according to the ratio of inner and outer sphere contributions to the reorganization energies obtained using polarizable solvent models. Specifically, we independently evaluated the intramolecular and environmental contributions to the energy gap,
\begin{equation}
    \Delta = \Delta_{\mathrm{mol}} + \Delta_{\mathrm{env}}
\end{equation}
A similar decomposition can be done for the reorganization energies sampled in the systems with polarizable solvent, 
\begin{equation}
    \lambda_{AB} = \lambda_{AB}^{(\mathrm{mol})} + \lambda_{AB}^{(\mathrm{env})}
\end{equation}
where the contribution from the cross-correlations between inner and outer sphere components was systematically verified to be negligible. From this, we can define the ratio,
\begin{equation}
    f_{\mathrm{mol}} = \frac{\lambda_{AB}^{(\mathrm{mol})}}{\lambda_{AB}}
\end{equation}
and the final spectral distributions were then calculated as
\begin{equation}
    \rho_{AB}(\omega) = f_{\mathrm{mol}}\rho_{AB}^{(\mathrm{mol})}(\omega) + (1-f_{\mathrm{mol}})\rho_{AB}^{(\mathrm{env})}(\omega)
\end{equation}
a weighted sum of the two contributions.

\begin{figure}[t]
    \centering
    \includegraphics[width=0.9\linewidth]{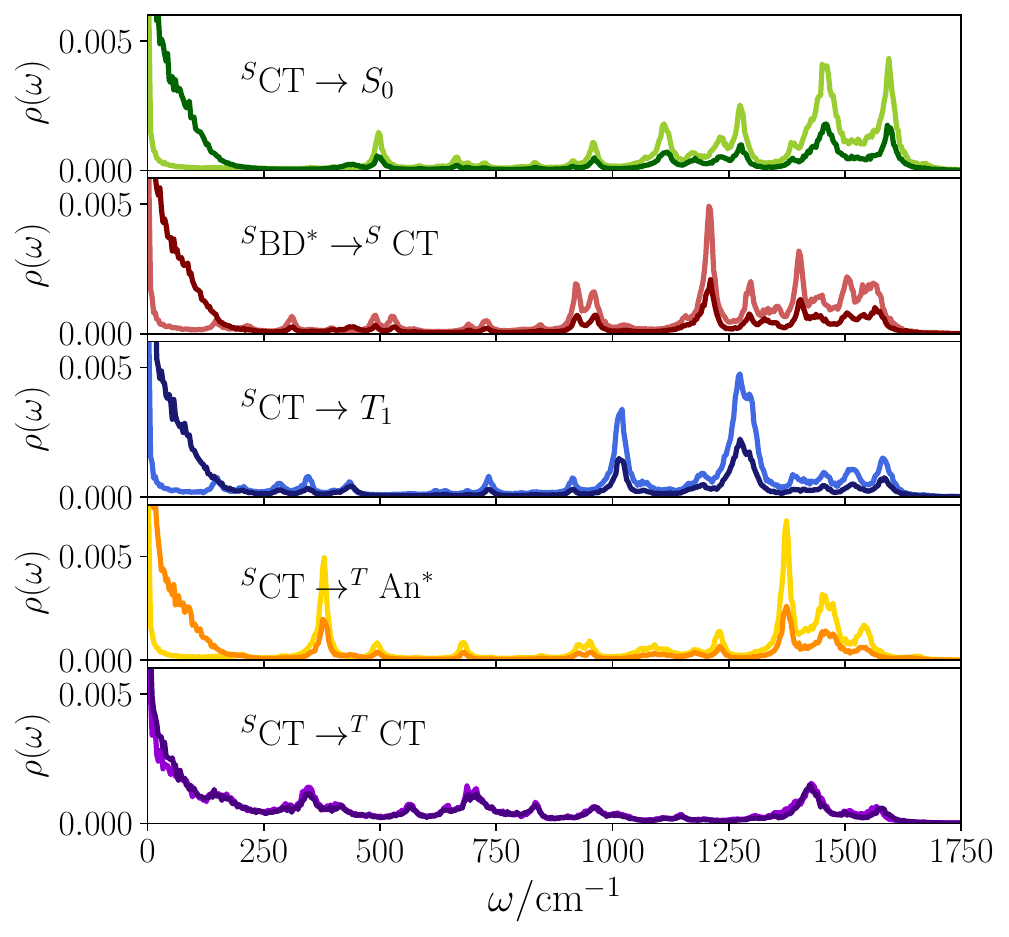}
    \caption{BoANTH spectral densities sampled in ACN (dark colors) and TOL (light colors) for a selection of electronic transitions}
    \label{fig:BoANTH_rho}
\end{figure}

\begin{figure}
    \centering
    \includegraphics[width=0.9\linewidth]{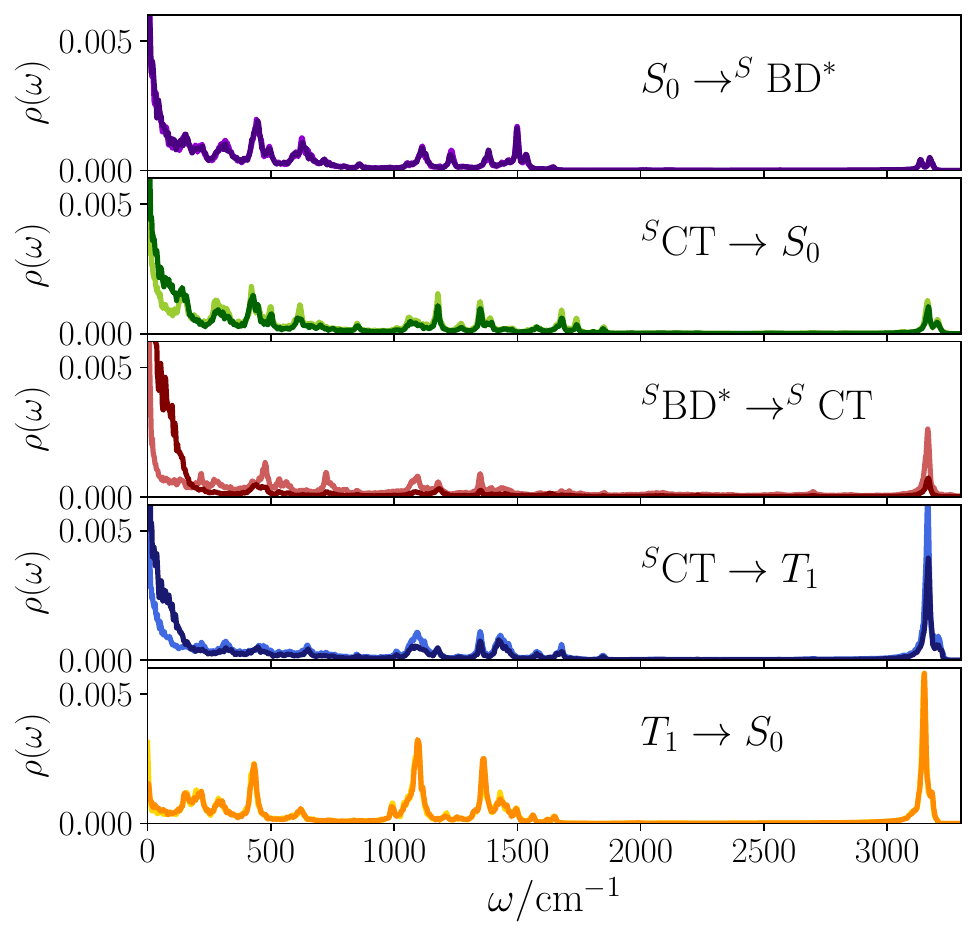}
    \caption{BoPTH spectral densities sampled in ACN (dark colors) and TOL (light colors) for a selection of electronic transitions}
    \label{fig:BiPTH_rho}
\end{figure}

The spectral distributions for a set of transitions for BoANTH and BoPTH in both ACN and TOL are shown in Figs. \ref{fig:BoANTH_rho} and \ref{fig:BiPTH_rho}, respectively. The most salient difference between the two solvents is that the low-frequency component of the spectral distribution, which is predominantly related to the outer-sphere solvent fluctuations, is substantially attenuated in TOL for transitions that involve charge transfer, implying a much smaller solvent contribution to the reorganization energy. This is consistent with predictions from dielectric continuum theory, where the outer-sphere component of the reorganization energy increases with solvent polarity. In transitions that do not involve significant charge rearrangement, the differences between solvents in the low-frequency region of the distributions are much smaller. 

The high frequency region of the spectral distributions is essentially unchanged between solvents, up to a scaling due to the diminished contribution from the low-frequency modes in TOL. 
Interestingly, BoPTH exhibits a feature at very high frequency ($\sim 3150 \ \wn$) in both solvents, which is most likely associated with an aromatic C-H stretching in the PTH fragment. To verify that this peak is not an artifact of the fitted forcefields, we calculated the gas-phase Huang-Rhys factors for the $S_0\to\sbo$ transition, which are shown in Appendix D. We confirmed that high-frequency normal modes are indeed coupled to this transition, giving rise to the observed feature. We note that this could also have implications for the degree to which nuclear quantum effects influence the non-adiabatic rates in this system. 


\subsection{Free energetics of excited state transitions}
The reorganization energies and driving forces for the various transitions can be extracted from the free energy surfaces of the energy gap,
\begin{equation}
    F_{A}(\Delta) = -\kB T\ln \langle\delta(\Delta-\Delta(\textbf{Q}))\rangle_A
\end{equation}
where we have absorbed an overall normalization constant that sets the zero of $F_{A}(\Delta)$. If $\Delta$ obeys Gaussian statistics, Marcus theory holds and $F(\Delta)$ will have parabolic form \cite{marcus1956theory, marcus1957theory, chandler1998electron}. However, this limiting behavior is not guaranteed for general systems, so it is important to construct free energy surfaces in a manner that is agnostic to whether the energy gap exactly obeys linear response. To this end, we sampled the energy gap in BoANTH through thermodynamic integration, whereby we generated biased ensembles that linearly interpolated between the Hamiltonians of the two states \cite{ferrario2006computer} 
\begin{equation}
    H_{\eta} = (1-\eta)H_{A} + \eta H_{B} = H_A + \eta \Delta 
\end{equation}
and obtained the full free energy surfaces from the energy gap distributions of these biased trajectories through histogram reweighting using MBAR \cite{shirts2008statistically}. An interval of 0.2 in the biasing parameter $\eta$ was found to produce sufficient overlap between the energy gap distributions of adjacent thermodynamic integration windows. The energy gap in each biasing window was sampled every 0.1 ps over trajectories of 200 ps in length. These simulations were once again performed at 298 K in the NVT ensemble with a Langevin thermostat in a system consisting of the photosensitizer and 1000 solvent molecules.

The free energy difference for a transition between $A$ and $B$ is given by
\begin{equation}
    e^{\beta \Delta F_{A,B}} = \left \langle e^{-\beta\Delta} \right \rangle_A = \int e^{-\beta [ F_A(\Delta)-\Delta]} d\Delta
\end{equation}
which is derivable from the free energy surface as a function of the energy gap. If linear response is exactly obeyed, $\Delta F_{A,B}$ may also be calculated as
\begin{equation}
    \label{LR_F}
    \Delta F_{A,B} = \frac{1}{2}\left(\langle\Delta\rangle_A+\langle\Delta\rangle_B\right)
\end{equation}
Analogously, the reorganization energy can be evaluated from a number of different expressions,
\begin{equation}
    \label{reorg_lr}
    \lambda_{AB} = \frac{1}{2}\left(\langle\Delta\rangle_B - \langle\Delta\rangle_A\right) = \beta \frac{\langle\delta\Delta^2\rangle_A}{2} = \beta \frac{\langle\delta\Delta^2\rangle_B}{2}
\end{equation}
We shall use the first equality in terms of difference in the means of $\Delta$ to calculate $\lambda_{AB}$ in our systems, but all definitions are equivalent if the environment is linearly responding. The corresponding reorganizations energies and driving forces are shown in Table \ref{tab:all_data}.

\begin{figure}[t]
    \centering
    \includegraphics[width=0.9\linewidth]{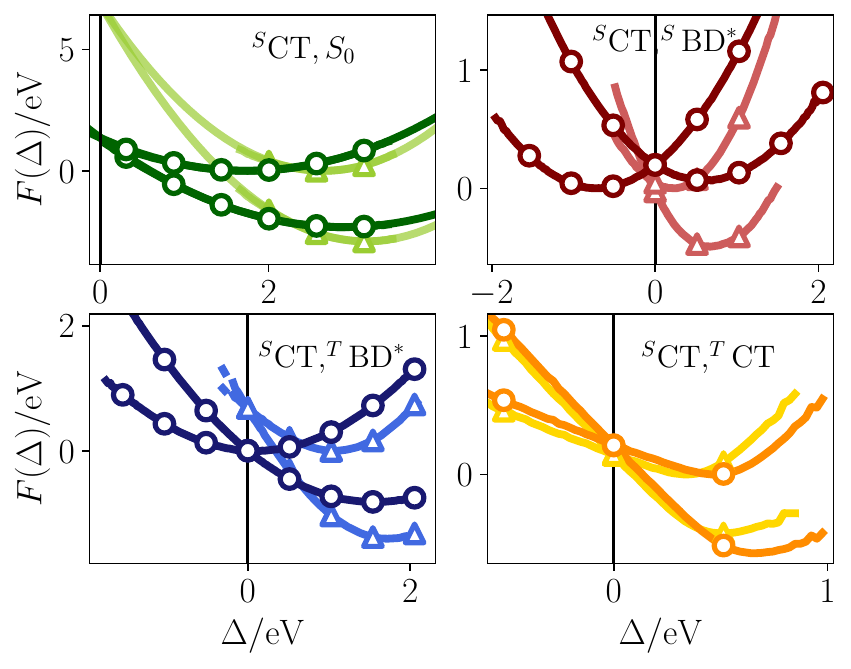}
    \caption{Free energy surfaces of select transitions in BoANTH dissolved in TOL(triangles), and ACN(circles)}
    \label{fig:BoANTH_FES}
\end{figure}

\begin{figure}[t]
    \centering
    \includegraphics[width=0.9\linewidth]{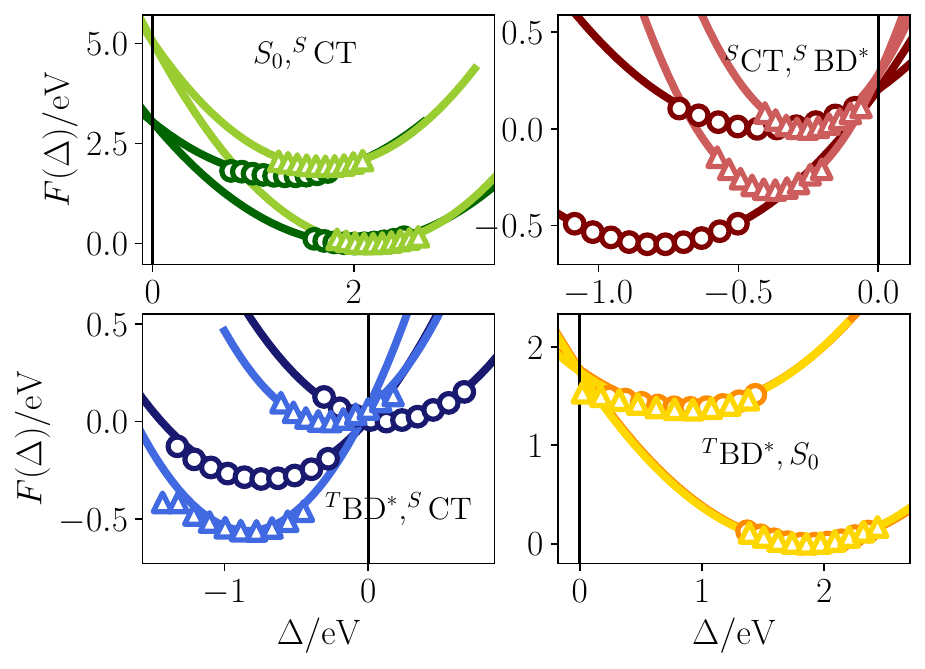}
    \caption{Free energy surfaces of select transitions in BoPTH dissolved in TOL(triangles), and ACN(circles). Markers correspond to the distribution directly sampled from simulation, while the solid lines are the corresponding Marcus free energy surfaces.}
    \label{fig:BoPTH_FES}
\end{figure}

A selection of free energy surfaces for BoANTH are shown in Fig. \ref{fig:BoANTH_FES}. In general, we find that the majority of the free energy surfaces are parabolic to a reasonably good approximation, consistent with the assumption of linear response. This is further confirmed by the fact that the various definitions of reorganization energy in Eq. \ref{reorg_lr} are generally in good agreement. Furthermore, we find that the values of $\Delta F_{A,B}$ computed using MBAR agree almost exactly with the  estimate in Eq. \ref{LR_F}, which only requires information from the unbiased ensembles. On the basis of this observation, we did not repeat the thermodynamic integration for BoPTH, and opted to rely on linear response to construct the free energy surfaces displayed in Fig. \ref{fig:BoPTH_FES}, where we see very good fit between the sampled distributions and the corresponding Marcus parabolic form. As expected, all of the reorganization energies of transitions involving charge transfer are systematically larger in ACN than in TOL. However, dielectric continuum theory predicts reorganization energies in ACN to be larger than those in TOL by a factor $\sim 8$, whereas we find that the reorganization energies in toluene are smaller only by a factor of 1.1-1.5. This is most likely due to the fact that the intramolecular nature of these transitions diminish the effect of solvent dielectric dispersion to some extent, rendering a pure continuum prediction inadequate.


The $\ct\to S_0$ reaction lies in the Marcus inverted regime in all the systems, but it is deeper in the inverted regime in TOL than ACN. This follows from the additional stability awarded to the $\ct$ state by the more polar solvent, and it implies that the non-radiative charge recombination pathway, a key competing mechanism in triplet state formation, will be enhanced in ACN with respect to TOL. This is especially relevant in the case of BoPTH, where charge separation is thermodynamically favorable in both solvents. In BoANTH, however, while charge separation is thermodynamically favorable in ACN, it becomes endoergic in TOL by nearly 0.5 $e$V, which significantly inhibits the formation of the charge transfer state to begin with. As for the SOCT-ISC transition from $\ct$ to $\tbo$ in BoANTH, it goes from nearly activationless in ACN to inverted in TOL with a driving force almost three times larger than the reorganization energy, once again suppressing the formation of triplet states in this system. In BoPTH, on the other hand, this process has quite similar reorganization energies and driving forces in the two solvents, and result in a process that is nearly activationless, albeit slightly in the inverted regime for the case of TOL.

\begin{table*}[t]
    \centering
    \begin{tabular}{cc||c|cccc|cccc}
        \hline
         A & B & &  & ACN & &  & & TOL & &\\
        \hline
        \hline
        BoANTH & & $V / \mathrm{cm}^{-1}$ & $\lambda_{AB} / e\mathrm{V}$ & $\Delta F_{A, B} / e\mathrm{V}$ & $k^{(\mathrm{NR})}_{A, B} / \mathrm{s}^{-1}$ & $k^{\mathrm{(R)}}_{A, B} / \mathrm{s}^{-1}$ & $\lambda_{AB} / e\mathrm{V}$ & $\Delta F_{A, B} / e\mathrm{V}$ & $k^{\mathrm{(NR)}}_{A, B} / \mathrm{s}^{-1}$ & $k^{\mathrm{(R)}}_{A, B} / \mathrm{s}^{-1}$ \\
        \hline
         $\sbo$ & $S_0$ & --- & 0.0845 & -2.385 & --- & 9.880$\times10^7$ & 0.0808 & -2.411 & --- & 1.026$\times10^8$ \\
         $^S \mathrm{CT}$ & $S_0$ & 1904 & 0.5953 & -2.320 & 1.466$\times 10^{8}$* & 3.289$\times10^{6}$ & 0.4213 & -2.904 & 4.783$\times 10^{3}$* & 9.054$\times10^6$\\
         $^S\mathrm{Bo}^*$ & $^S\mathrm{CT}$ & 99 & 0.6657 & -0.065 & 4.452$\times 10^{10}$ & --- & 0.4631 & 0.493 & 1.129$\times10^4$ & ---  \\
         $^S\mathrm{CT}$ & $^T\mathrm{Bo}^*$ & 0.79 & 0.7547 & -0.817 & 1.289$\times 10^{8}$ & --- & 0.5387 & -1.400 & 1.302$\times 10^{6}$ & ---  \\
         $^S\mathrm{CT}$ & $^T\mathrm{ANTH}^*$ & 0.63 & 0.5992 & -0.462 & 9.590$\times 10^{7}$ & --- & 0.4286 & -1.022 & 4.591$\times10^{6}$ & ---  \\
         $^S\mathrm{CT}$ & $^T\mathrm{CT}$ & 0.21 & 0.1109 & -0.567 & 3.708$\times 10^{5}$ & --- & 0.1021 & -0.425 & 7.478$\times 10^{5}$ & ---  \\
        \hline
        \hline
        BoPTH & & $V / \mathrm{cm}^{-1}$ & $\lambda_{AB} / e\mathrm{V}$ & $\Delta F_{A, B} / e\mathrm{V}$ & $k^{(\mathrm{NR})}_{A, B} / \mathrm{s}^{-1}$ & $k^{\mathrm{(R)}}_{A, B} / \mathrm{s}^{-1}$ & $\lambda_{AB} / e\mathrm{V}$ & $\Delta F_{A, B} / e\mathrm{V}$ & $k^{\mathrm{(NR)}}_{A, B} / \mathrm{s}^{-1}$ & $k^{\mathrm{(R)}}_{A, B} / \mathrm{s}^{-1}$ \\
        \hline
         $\sbo$ & $S_0$ &  --- & 0.1880 & -2.2597 & --- & 9.535$\times 10^{7}$ & 0.1821 & -2.2343 & --- & 9.291$\times 10^{7}$ \\
         $^S \mathrm{CT}$ & $S_0$ & 5141.90 & 0.3408 & -1.6795 & 1.968$\times 10^{7}$* & 1.516$\times 10^{6}$ & 0.2865 & -1.9307 & 2.115$\times 10^{6}$* & 2.779$\times 10^{6}$ \\
         $^S\mathrm{Bo}^*$ & $^S\mathrm{CT}$ & 114.27 & 0.1988 & -0.5961 & 2.414$\times 10^{11}$ & --- & 0.1367 & -0.3161 & 1.116$\times 10^{12}$ & ---  \\
         $^S\mathrm{CT}$ & $^T\mathrm{Bo}^*$ & 0.72 & 0.4377 & -0.2942 & 1.485$\times 10^{8}$ & --- & 0.3631 & -0.5601 & 6.527$\times 10^{7}$ & ---  \\
         $^T\mathrm{Bo}^*$ & $S_0$ & 0.30 & 0.5014 & -1.3793 & 3.038$\times 10^{5}$ & --- & 0.4920 & -1.3619 & 3.182$\times 10^{5}$ & ---  \\
        \hline
    \end{tabular}
    \caption{Diabatic couplings, reorganization energies, driving forces and nonadiabtic rates of all the relevant transitions between excited states in BoANTH and BoPTH dissolved in ACN and TOL. \footnotesize{* Non-adiabatic rate corrected by the OGR method}}
    \label{tab:all_data}
\end{table*}

\section{Lifetimes and triplet yields}\label{sec4}
The non-radiative rates, shown in Table \ref{tab:all_data}, were calculated from the time-integral of the flux-correlation function evaluated at the free energy of the transition. 
Figures of the rates as a function of $\Delta F$ from Eq. \ref{nonrad_rate} are shown in Appendix E. These rate profiles clearly exemplify the additional level of insight garnered by explicit solvation models. Beyond changes in the driving force induced by the solvent, our method captures contributions to the spectral density from the full frequency-dependence of the solvent dielectric dispersion, thus capturing static and dynamical solvent effects well beyond a simple continuum theory.
Because of their large diabatic couplings, the $\ct\to S_0$ rates were corrected following the OGR procedure\cite{fay2024extending}. In essence, this correction relies on computing the rate on a new diabatic basis $\{\ket{+}, \ket{-}\}$, with the transformation parametrized by a rotation angle $\theta$,
\begin{align}
    \ket{+} &= \ket{A}\cos(\theta) + \ket{B}\sin(\theta) \\
    \ket{-} &= -\ket{A}\sin(\theta) + \ket{B}\cos(\theta) 
\end{align}
This angle is chosen such that the diabatic coupling is minimized, resulting in the following equation for the optimal value of $\theta$,
\begin{align}
    \tan(2\theta_A) &= \frac{2V}{\langle\Delta\rangle}_A \\
    \tan(2\theta_B) &= \frac{2V}{\langle\Delta\rangle}_B
\end{align}
The final rate is then chosen as the minimal value within the range in the $[\theta_A,\theta_B]$ interval,
\begin{equation}
    k_{\mathrm{OGR}} = \min_{\theta\in[\theta_A,\theta_B]} k_\theta
\end{equation}
where $k_\theta$ is the rate evaluated in the diabatic basis at fixed $\theta$.
In BoANTH, we find that the OGR correction reduces the non-radiative charge recombination rate by a factor $\sim 4$, but in BoPTH we observe a dramatic suppression of this rate, by roughly 3 orders of magnitude. This difference can be attributed to the diabatic coupling being nearly three times larger in BoPTH than in BoANTH. While the Golden Rule rate can become exponentially inaccurate at large coupling, the OGR correction has been systematically shown to produce reliable and accurate results even in very strong coupling regimes,\cite{fay2024extending} making it crucial for the correct description of these systems while remaining at the lowest order in perturbation theory. 
For all other transitions, the OGR rate differs negligibly from the Fermi's Golden rule estimate. Given that this correction will only begin to deviate from the Golden rule as the diabatic coupling increases, this indicates that all the other electronic transitions are well within the range of validity of weak coupling perturbation theory.

For BoANTH in ACN, we find results consistent with previous studies,\cite{fay2024unraveling} with a very fast charge separation from $\sbo$ into the $\ct$ state with a rate of $4.452\times10^{10} \ \mathrm{s}^{-1}$, and subsequently facile formation of the $\tanth$ and $\tbo$ triplet states with rates on the order of $10^{8} \ \mathrm{s}^{-1}$. The symmetry-forbidden transition into $\tct$ has a naturally smaller rate of $3.708\times 10^{5} \ \mathrm{s}^{-1}$, but it is still a non-negligible contribution to the accumulation of triplet population. In TOL, these trends are essentially reversed. The large thermodynamic cost of charge separation results in an exceedingly small rate of only $\sim 10^4 \ \mathrm{s}^{-1}$. Furthermore, the intersystem crossing rates are also considerably smaller in this solvent. Since $\ct$ is significantly less stable, these processes go from nearly activationless in ACN to inverted in TOL, resulting in rates that are $\sim$ 2 orders of magnitude smaller than in the high dielectric medium.

Triplet state formation is suppressed by both radiative and non-radiative decay to $S_0$. To evaluate the radiative rates of spontaneous emission to the ground-state, $k^{(\mathrm{R})}_{A\to S_0}$ from state $A$, we use
\begin{equation}
    k^{(\mathrm{R})}_{A, S_0} = \frac{|\mu_{A,S_0}|^2}{3\pi\epsilon_0\hbar c_0^3} \int_0^\infty F_{A,S_0}(\omega) \omega^3 d\omega
\end{equation}
where  $\epsilon_0$ and $c_0$ are the vacuum permittivity and speed of light, respectively, and $\mu_{A,S_0}$ is the transition dipole between $S_0$ and state $A$. The normalized emission lineshape, $F_{A,B}(\omega)$ is
\begin{equation}
    F_{A,B}(\omega) = \frac{1}{2\pi}\int_{-\infty}^{\infty} e^{-i\omega t} c^*_{A,B}(t) dt
\end{equation}
and $c^*_{A,B}(t)$ is the complex conjugate of $c_{A,B}(t)$. For BoANTH in ACN, the dominant channel for relaxation to the ground state is non-radiative charge recombination, with a rate of $1.466\times 10^{8} \ \mathrm{s}^{-1}$.  For BoANTH in TOL, charge recombination is even deeper in the inverted regime, resulting in a dramatically smaller rate of $4.783\times10^{3} \ \mathrm{s}^{-1}$, the excited state dynamics are overwhelmingly dominated by radiative decay to $S_0$ from $\sbo$ before the charge transfer state is formed. Additionally, due to the crossover to Marcus inverted kinetics, the rates of intersystem crossing themselves are also smaller in TOL than in ACN by as much as 2 orders of magnitude. 

In BoPTH, the charge separation rates in both solvents are very high, on the order of $10^{11} - 10^{12}\ \mathrm{s}^{-1}$. As a consequence, $\sbo$ and $\ct$ will rapidly achieve a pre-equilibrium upon photoexcitation, with the vast majority of the population concentrated in the $\ct$ state, thus allowing intersystem crossing. As mentioned previously, a crucial difference between the two solvents lies in the non-radiative charge recombination process to $S_0$, where the exoergicity in TOL exceeds that in ACN by $\sim 0.25 e$V, resulting in a rate that is smaller by roughly one order of magnitude. The SOCT-ISC rate is about twice as large in ACN, once again likely due to the fact that this transition lies slightly deeper in the inverted regime for TOL. 

The triplet yield, $\Phi_T$ of each system was calculated assuming a pre-equilibrium is attained between the $\sbo$ and $\ct$ states, allowing us to define a global charge recombination rate as 
\begin{multline}
    \label{kCReff}
    k_{\mathrm{CR}} = p_{\sbo}k^{(R)}_{\sbo, S_0}+ \\
    p_{\ct}\left(k^{(R)}_{\ct, S_0}+k^{(NR)}_{\ct, S_0}+k^{(NR)}_{\ct, T}\right)
\end{multline}
where the thermal populations of singlet and charge transfer states are
\begin{equation}
    p_{\sbo} = \frac{1}{1+e^{-\beta \Delta F_{
^S\mathrm{Bo}^*,^S\mathrm{CT} }
    }} \  , \ p_{\ct} = 1- p_{\sbo}
\end{equation}
and $k^{(NR)}_{\ct, T}$ is the sum of all the rates from $\ct$ to the triplet manifold. With these definitions, $\Phi_T$ can be calculated as
\begin{equation}
    \Phi_T = \frac{\ p_{\ct}k^{(NR)}_{\ct, T}}{k_{\mathrm{CR}}}
\end{equation}
just the ratio between the transition from the transition into the triplet manifold from the charge transfer state, relative to the  global charge recombination rate.
Additionally, the fluorescence yield, $\Phi_F$ was computed as,
\begin{equation}
    \Phi_F = \frac{p_{\sbo}k^{(R)}_{\sbo, S_0}+p_{\ct}k^{(R)}_{\ct, S_0}}{k_{\mathrm{CR}}}
\end{equation}
and $\phi_{\mathrm{NRT}}$, the efficiency of triplet formation due solely to non-radiative processes, is given as
\begin{equation}
    \phi_{\mathrm{NRT}} = \frac{\Phi_T}{1-\Phi_F}
\end{equation}
A comparison of the triplet yields, fluorescence yields, and non-radiative efficiency across all systems is shown in Table \ref{tab:rate_data}.

The pre-equilibirum assumption underlying Eq. \ref{kCReff} was tested by explicitly propagating a kinetic model through a master equation where the rate matrix  is constructed from the calculated rates, and the initial conditions were chosen to represent a vertical photoexcitation populating the $\sbo$ state.
Given that the rates of decay to the ground state from the triplet manifold are by far the smallest, these decay channels were not included in the kinetic model, such that both $S_0$ and $\tbo$ act as sinks. This allows to evaluate the triplet yield as the steady-state population of the $\tbo$ state. 
In practice, we find that the triplet yields calculated by direct propagation to steady-state are essentially identical to the ones derived from the pre-equilibrium approximation even for cases, such as BoANTH in toluene, where a pre-equilibrium assumption would appear to be invalid.
The propagation of the dynamics as a classical kinetic model neglects possible effects from interference in the quantum dynamics. It is an approximation justified on the grounds that the charge separation and recombination processes take place over fairly disparate timescales.

\begin{table}[b!]
    \centering
    \begin{tabular}{c||ccc|ccc}
        \hline
          & &ACN & & &TOL& \\
          \hline
          \hline
          & $\Phi_F$ & $\Phi_T$ & $\phi_\mathrm{NRT}$ & $\Phi_F$ & $\Phi_T$ & $\phi_\mathrm{NRT}$ \\
          \hline
        BoANTH & 0.031 & 0.587 & 0.606 & $\sim$ 1 & $\sim$ 0 & 0.990 \\
        BoPTH & 0.013 & 0.875 & 0.886 & 0.030 & 0.930 & 0.959\\
        \hline
    \end{tabular}
    \caption{Comparison of triplet yields across sensitizers and solvents.}
    \label{tab:rate_data}
\end{table}

Qualitatively, we find agreement with the basic experimental observations that motivated this study. Namely, we find larger triplet yields in BoANTH embedded in the more polar ACN solvent, and the opposite for BoPTH. It is, however, readily apparent that our results do not provide quantitative predictions of the kinetic data. This is especially true of BoANTH in TOL where the calculated triplet yield is vanishingly small, whereas experiment reports a singlet oxygen yield of 0.38, about $50\%$ smaller than in ACN, but still significant. The reason behind the discrepancy can be attributed almost entirely to the slow charge separation rate completely inhibiting the formation of $\ct$. The non-radiative efficiency in TOL is significantly larger, reflecting that in the absence of radiative decay, the slower non-radiative $\ct\to S_0$ transition allows for the formation of some triplet states, despite the SOCT-ISC rate being smaller than in ACN. 

In BoPTH, it is clear that our calculated rates overestimate of the extent of triplet formation, with a value of 0.930 in TOL being $\sim 27 \%$ larger than the experimental singlet oxygen yield. While it is true that the singlet oxygen yield corresponds to a strict lower bound to $\Phi_T$, we observe only slightly a smaller triplet yield in ACN, which is experimentally reported to form triplet states in negligible amounts. The smaller value of $\Phi_T$ in ACN can be directly linked to the enhanced rate of non-radiative charge recombination that results from dielectric stabilization of the $\ct$ state. The calculated fluorescence yields are in considerably better agreement with those inferred experimentally, these being 0.004 in ACN and 0.010 in TOL\cite{uddin2023twist}, with our results being similarly small and the ratio in fluorescence yields between TOL and ACN also being roughly 2. This indicates that our description successfully predicts that decay to the ground state is dominated by non-radiative channels. 

\begin{figure}
    \centering
    \includegraphics[width=0.7\linewidth]{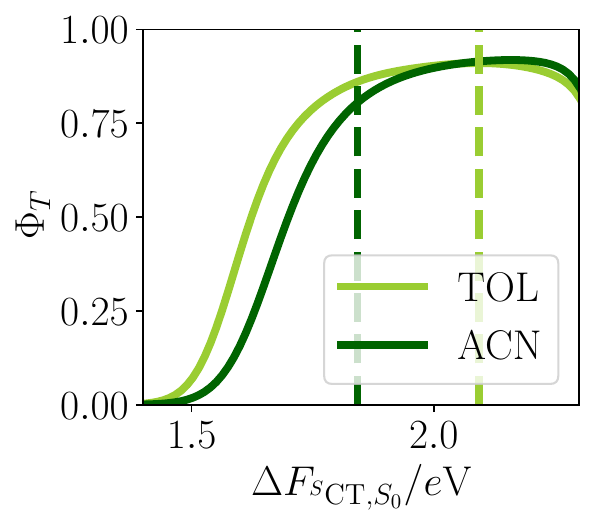}
    \caption{Triplet yield in BoPTH as a function of the energy of the $\ct$ state, while keeping all other state energies fixed. Dashed lines correspond to the calculated free energy of this state in either solvent.}
    \label{fig:yield_scan}
\end{figure}

We believe that the most relevant source of error in our calculations are inaccuracies in the electronic excitation energies of all states, especially that of the $\ct$ state. The triplet yield can be exponentially sensitive to the free energy differences between states, and small variations easily lead to quite dramatic changes in $\Phi_T$. Figure \ref{fig:yield_scan} shows the dependence of the triplet yield with respect to the free energy of the $\ct$ state, which was varied while keeping all other state energies constant. This analysis is especially revealing when considering that the typical error of EOM-CCSD is of 0.1 $e$V - 0.3 $e$V, and in charge-transfer excitations, this error has been reported to be systematically positive. Looking at the trends in Fig. \ref{fig:yield_scan}, it can be seen that a uniform redshift of this magnitude in $\Delta F_{\ct,S_0}$ would result in a significant reduction of triplet yield in ACN, while that in TOL would remain fairly large, leading to values that are in much better agreement with experiment. Fig. \ref{fig:yield_scan} ultimately reveals that the reliability of electronic excitation energies remains one of the biggest challenges for the quantitative description of excited state dynamics. However, it also confirms that the \textit{qualitative} solvent effects predicted by our framework are robust to moderate changes in the electronic excitations energies. 

\section{Conclusion}\label{sec5}
In this study, we have investigated the photophysics of two different heavy atom-free triplet sensitizers dissolved in both ACN and TOL. The spin-boson mapping that we constructed allowed us to probe these systems and sample spectral densities with atomistic detail. Our model confirmed that differences in dielectric stabilization of the charge-transfer state are the key ingredient driving the differences in activity of these two molecules in environments of high and low polarity. This observation leads to the qualitatively correct predictions in the trends of triplet yield across solvents.

Despite providing a good qualitative description and important molecular insight, our calculations fail to find good agreement with the experimental values of triplet yield. A discussion of the potential sources of this discrepancy is in order. Firstly, the rates depend exponentially on the driving force, so an error of only a few $\kB T$ in $\Delta F$ can easily change the rate by an order of magnitude. It follows that, despite using relatively high-level methods to compute the electronic gaps, and empirically correcting them based on the experimental absorption peak, errors in the electronic energies are likely to be the leading cause for the inability of our results to reproduce the experimental triplet yields. As discussed above, potential errors in EOM-CCSD charge transfer excitation energies could account for the inaccuracies observed in $\Phi_T$ for the BoPTH molecule, and they can also lead to a positive bias in the energy of charge separation $\sbo\to\ct$, which might be the reason why the charge separation process was  found to be prohibitively unfavorable in BoANTH dissolved in toluene.

It is also possible that, despite not finding them in our calculations, there are additional triplet states in BoPTH that are energetically-accessible and contribute to the total SOCT-ISC rate. Another potential source of error comes from the assumptions made when constructing the effective model for the $\ct$ state in BoPTH. While our calculations indicate that a planar conformation is not stable in solution, if the puckered-to-planar conformational change does play a role in the charge separation and recombination dynamics, correctly capturing this mechanism is beyond the scope of the spin-boson mapping, which relies on moderately low-amplitude displacements in the bath degrees of freedom, allowing them to be described as locally harmonic.  

\section*{Acknowledgments}
This work was supported by the Condensed Phase
and Interfacial Molecular Science Program (CPIMS) of
the U.S. Department of Energy under contract no. DEAC02-05CH11231.

\section*{Data Availability}
Forcefield files for solvents and all electronic states of both photosensitizers, as well as initial configurations and example OpenMM scripts to carry out energy gap calculations can be found on GitHub\cite{SuppData}.  Other data supporting the findings of this study are available from the corresponding author
upon reasonable request.

\appendix

\subsection*{Appendix A}
In BoANTH, the donor-localized ($\tanth$) and charge-transfer ($\tct$) triplet states have been invoked as important contributors to triplet state formation. The $\tanth$ state in particular has been found in this work to have an SOCT-ISC rate almost as large as the $\ct\to\tbo$ transition, while the $\tct$ state plays only a minor role. We find, however, that for BoPTH these states do not contribute appreciably to triplet formation. 

\begin{figure}[h]
    \centering
    \includegraphics[width=\linewidth]{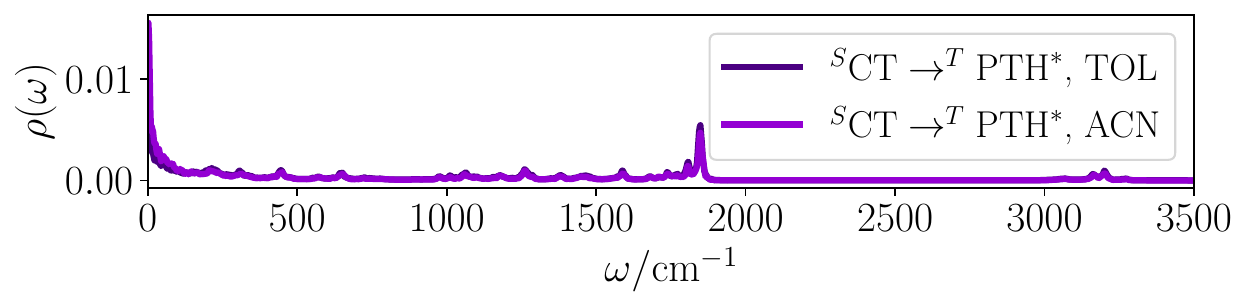}
    \caption{Spectral distribution for the $^S\mathrm{CT} \to ^T\mathrm{PTH}^*$ transition in both ACN and TOL}
    \label{fig:sct_tpth_specdens}
\end{figure}

The $^T\mathrm{PTH}^*$ state was identified and optimized using the same procedure and level of theory as all the other states. Similarly, the excited state force field used to sample free energies and spectral densities (shown in Fig. \ref{fig:sct_tpth_specdens} for the $\ct\to^T\mathrm{PTH}^*$ transition) was constructed following the same procedure as described in the main text. While we find that the gas-phase electronic energy of this state is moderately low (2.27 eV above the $S_0$ minimum), the SOCT-ISC free energy change is slightly exoergic in both solvents due to the dielectric stabilization of the charge-transfer state. We also find, both in the electronic structure calculations and the molecular simulations, that this transition exhibits quite large values of inner-sphere reorganization energy, in the order of 1.0 eV. The exorergicity of the transition alongside the large values of reorganization energy results in extremely small rate values for the SOCT-ISC rate, shown in Table III, thus contributing negligibly to the triplet yield.  

\begin{tabular}{cc|c|ccc}
        \hline
         A & B & $V / \mathrm{cm}^{-1}$ & $\lambda_{AB} / e\mathrm{V}$ & $\Delta F_{A, B} / e\mathrm{V}$ & $k^{(\mathrm{NR})}_{A, B} / \mathrm{s}^{-1}$ \\
        \hline
        \hline
         & &  &  & ACN & \\
        \hline
        $^S\mathrm{CT}$ & $^T\mathrm{PTH}^*$ & 0.236 & 1.965 & +0.321 & 2.25$\times10^{-3}$  \\
        \hline
         &  & &   & TOL & \\
        \hline
        $^S\mathrm{CT}$ & $^T\mathrm{PTH}^*$ & 0.236& 1.772 & +0.226 & 0.235 \\
        \hline
\end{tabular}
\begin{center}
    Table III: SOCT-ISC rates into the $^T\mathrm{PTH}^*$ state of BoPTH in ACN and TOL
\end{center}

We expect that the contribution of the  $\tct$ state  to the triplet yield is also likely negligible for two reasons. Firstly, this transition is strongly disfavored by El-Sayed's rule. Preliminary spin-orbit coupling calculations confirm that the $\ct$ and $\tct$ states are weakly coupled, with couplings of only $\sim 0.2 \wn$, which are over 3 times smaller than the $\ct\to\tbo$ transition. Secondly, our BoANTH results show that transitions into this state account for only 1.6 $\%$ of the total triplet yield in ACN and 2.4 $\%$ in TOL. Given that the SOCT-ISC rates that we have obtained in BoPTH are close in magnitude to those of BoANTH, and the energetics in the electronic structure calculations are similar, it is very likely that the contribution from this $\tct$ state is equally small in BoPTH. As an additional test, we constructed a minimal model for the $^T\mathrm{CT}$ state consisting of identical force constants to the $^S\mathrm{CT}$ state, while changing the atomic partial charges to reflect subtle differences in the charge rearrangement of the two excitations. Simulations with this model system show that the electrostatic contribution to the thermodynamic driving force of the $^S\mathrm{CT}\to ^T\mathrm{CT}$ transition is small but exoergic in both solvents, of +0.10 eV in TOL and +0.18 eV in ACN.

\subsection*{Appendix B}
Interstate couplings evaluated at the equilibrium geometry of the relevant electronic states for both dyads are presented in Table IV. We find that the Condon approximation is reasonable in general, and it is especially appropriate for SOCT-ISC transitions due to the small changes in the spin-orbit coupling as function of changes in geometry. This is consistent with previous and more recent studies,\cite{garzonramirez2026symmetrygovernsdihedralangle} confirming that non-Condon effects are not substantial in SOC transitions within this class of molecules, where the equilibrium dihedral angle of the dyad is not significantly changed between states.\cite{kosaka2024molecular} \\

\begin{tabular}{c|c|c|c}
    \hline
    \multicolumn{2}{c|}{States} & \multicolumn{2}{|c}{BoPTH} \\
    \hline
    A & B & $V_{AB}$@A geometry & $V_{AB}$@B geometry \\
    \hline
    $S_1$ & $^S\mathrm{CT}$ &  121 cm$^{-1}$ & 161 cm$^{-1}$  \\
    $S_0$ & $^S\mathrm{CT}$ & 2053 cm$^{-1}$ &  6975 cm$^{-1}$  \\
    $T_1$ & $S_0$ & 0.271 cm$^{-1}$ & 0.321 cm$^{-1}$  \\
    $T_1$ & $^S\mathrm{CT}$ & 0.771 cm$^{-1}$ & 0.541 cm$^{-1}$  \\
    $^T\mathrm{D}^*$ & $^S\mathrm{CT}$ & 0.244 cm$^{-1}$ & 0.228 cm$^{-1}$\\
    $^T\mathrm{CT}$ & $^S\mathrm{CT}$ & 0.349 cm$^{-1}$ & 0.228 cm$^{-1}$ \\
    \hline
    \multicolumn{2}{c|}{States} & \multicolumn{2}{|c}{BoANTH}\\
    \hline
    A & B & $V_{AB}$@A geometry & $V_{AB}$@B geometry \\
    \hline
    $S_1$ & $^S\mathrm{CT}$ & 64 cm$^{-1}$ & 125 cm$^{-1}$ \\
    $S_0$ & $^S\mathrm{CT}$ & 865 cm$^{-1}$ & 2550 cm$^{-1}$ \\
    $T_1$ & $S_0$ & 0.170 cm$^{-1}$ & 0.210 cm$^{-1}$ \\
    $T_1$ & $^S\mathrm{CT}$ & 0.790 cm$^{-1}$ & 0.800 cm$^{-1}$ \\
    $^T\mathrm{D}^*$ & $^S\mathrm{CT}$ & 0.620 cm$^{-1}$ & 0.640 cm$^{-1}$ \\
    $^T\mathrm{CT}$ & $^S\mathrm{CT}$ & 0.280 cm$^{-1}$ & 0.110 cm$^{-1}$ \\
    \hline
\end{tabular}
\begin{center}
    Table IV: Couplings evaluated at the equilibrium geometries of the two electronic states involved in the transition
\end{center}

\section*{Appendix C}
\setcounter{equation}{0}
\renewcommand{\theequation}{C\arabic{equation}}
The specific functional form of the forcefield we chose for the parametrization \cite{jorgensen1996development, dodda2017ligpargen} consists of ab some of contributions. This includes harmonic local stretching $U_{\mathrm{bond}}$ potential 
\begin{align}
    U_\mathrm{bond} &= \sum_{n\in \mathrm{bonds}} k^{\mathrm{b}}_n(r_n-r_{n,eq})^2
\end{align}
with $k^\mathrm{b}_n$ being the force constant for the $n$th bond, $r_n$ the displacement of the bond and $r_{n,eq}$ its rest length. It includes a bending $U_{\mathrm{angle}}$ potential,
\begin{align}
    U_\mathrm{angle} &= \sum_{n\in \mathrm{angles}} k^{\mathrm{a}}_n(\theta_n-\theta_{n,eq})^2
\end{align}
that acts on the angle between triplets of bonded atoms, where $k^\mathrm{a}_n$ being the force constant for the $n$th angle, $\theta_n$ the instantaneous angle and $\theta_{n,eq}$ its equilibrium angle. The potential also allows for interactions between bonds that share an atom $U_{\mathrm{bond-bond}}$, 
\begin{align}
    U_\mathrm{bond-bond} &= \sum_{\langle n,m\rangle} k^\mathrm{b}_{nm} (r_n-r_{n,eq})(r_m-r_{m,eq})
\end{align}
with force constant $k^\mathrm{b}_{nm}$ and between bonds and angles
$U_{\mathrm{bond-angle}}$, 
\begin{align}
    U_\mathrm{bond-angle} &= \sum_{\langle n,m\rangle} k^{b,a}_{nm} (r_n-r_{n,eq})(\theta_m-\theta_{m,eq})
\end{align}
with force constant $k^{b,a}$. We describe potential due to proper and improper dihedrals as $U_{\mathrm{torsion}}$,
\begin{align}
    U_\mathrm{torsion} = \sum_n  &\frac{k_n^{\mathrm{t},(1)}}{2} \left[1+\cos(\phi_n-f_{n,1})\right] +  \\
    &\frac{k_n^{\mathrm{t},(2)}}{2} \left[1-\cos(2\phi_n-f_{n,2})\right] + \\
    &  \frac{k_n^{\mathrm{t},(3)}}{2} \left[1+\cos(3\phi_n-f_{n,3})\right] 
 \end{align} 
 with constants $k_n^{\mathrm{t},(i)}$ and dihedral angle $\phi_n$ and phases $f_{n,i}$.
 Finally we include non-bonded interactions for the Lennard-Jones $U_{\mathrm{LJ}}$ and Coulomb $U_{\mathrm{Coul}}$ interactions. Their explicit forms depend on separation distance $r_{ij}$ as   
 \begin{align}
    U_\mathrm{LJ} &= \sum_{i<j} 4\epsilon_{ij}\left[\left(\frac{\sigma_{ij}}{r_{ij}}\right)^{12}-\left(\frac{\sigma_{ij}}{r_{ij}}\right)^6\right]\\
\end{align}
with energy scale $\epsilon_{ij}$ and size $\sigma_{ij}$ and
\begin{align}
    U_\mathrm{Coul} &= \sum_{i<j} \frac{q_iq_j}{4\pi\epsilon_0r_{ij}}
\end{align}
with charge $q_i$.
Lennard-Jones parameters were assumed to be the same for each electronic state and are taken from the OPLS-AA forcefield. The long-range forces are ignored for 1-2 and 1-3 bonded atoms and scaled by 0.5 for 1-4 bonded atoms.

\begin{figure}[t]
    \centering
    \includegraphics[width=0.9\linewidth]{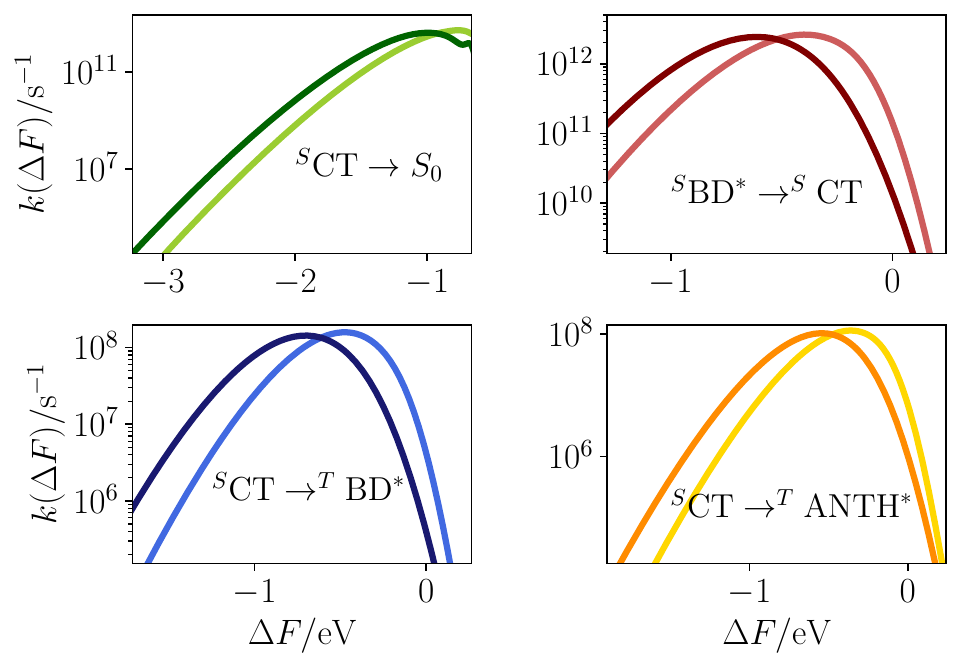}
    \caption{BoANTH Rates in ACN(dark colors) and TOL(light colors)}
    \label{fig:BoANTH_rates}
\end{figure}
\begin{figure}
    \centering
    \includegraphics[width=0.9\linewidth]{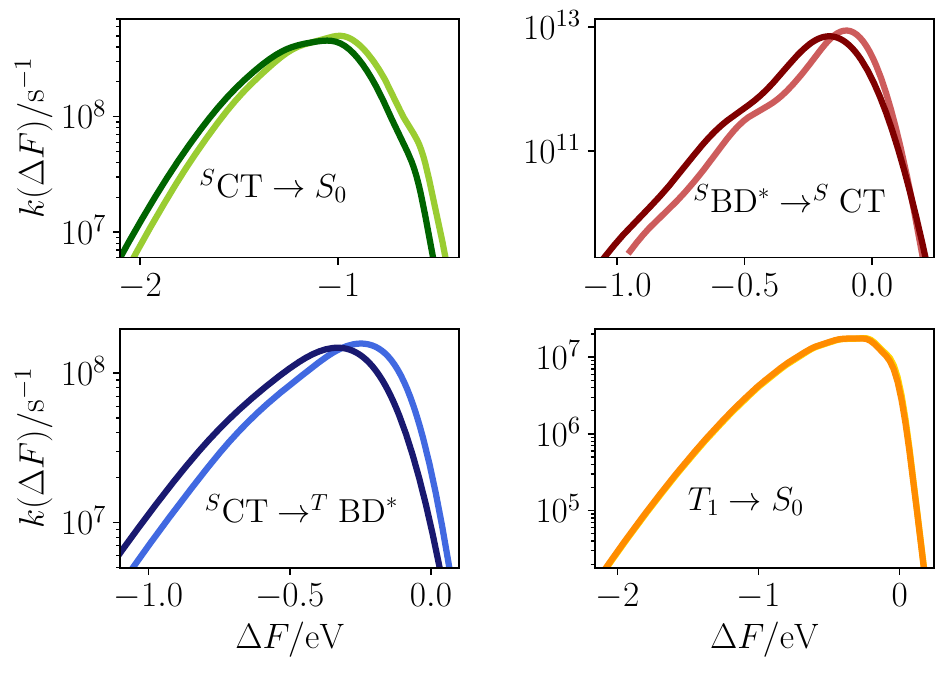}
    \caption{BoPTH Rates in ACN(dark colors) and TOL(light colors)}
    \label{fig:BoPTH_rates}
\end{figure}

The solvent models were constructed in a similar way as the forcefields of the molecular photosensitizers. The gas phase equilibrium geometries and Hessians of both ACN and TOL were calculated at the RI-MP2/cc-pVQZ level of theory. Vibrational forcefield parameters were fit to the quantum mechanical Hessians through the procedure described in the main text. Atomic partial charges were assigned as a linear combination of gas phase and CPCM CHELPG charges, chosen to reproduce the bulk density and dielectric constant of the solvents. 

For the polarizable solvent models, the Lennard Jones $\sigma$ and $\epsilon$ parameters were also adjusted to reproduce the bulk density and vaporization enthalpy, respectively. The Drude parameters, atomic polarizabilities and Thole damping parameters for TOL were taken from the CHARMM polarizable forcefield \cite{lopes2007polarizable}, and it was verified that the resulting combination reproduced not only the static and optical dielectric constants of TOL, but also the Debye relaxation times of the solvent \cite{santarelli1967overlapping}. The Drude parameters for the ACN model were fit from quantum chemistry polarizablity calculations, as described in our previous study \cite{fay2024unraveling}. For this solvent, the optical dielectric constant and Debye relaxation time were also verified to be in god agreement with experimental measurements.  

\subsection*{Appendix D}
\begin{figure}[t]
    \centering
    \includegraphics[width=0.9\linewidth]{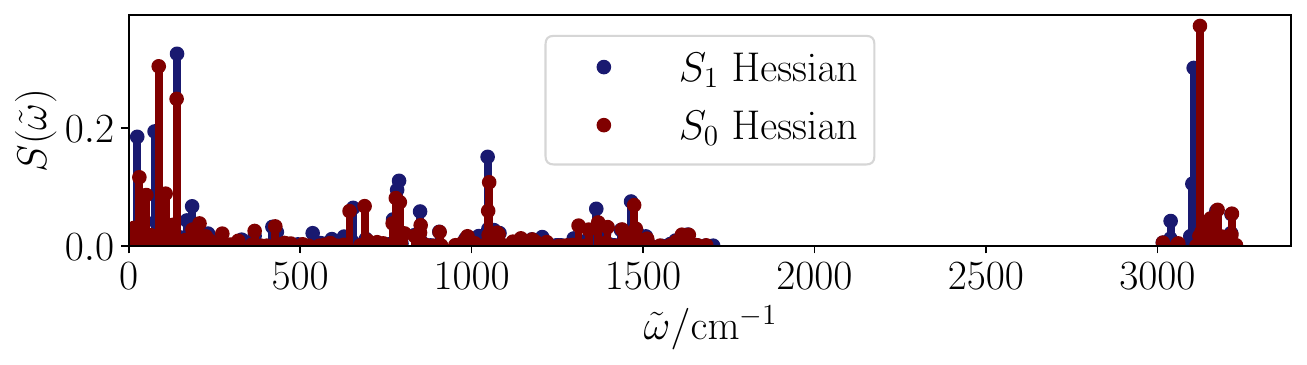}
    \caption{Gas-phase Huang-Rhys factors for the transition between $S_0$ and $\sbo$}
    \label{fig:Huang-Rhys}
\end{figure}

\setcounter{equation}{0}
\renewcommand{\theequation}{D\arabic{equation}}
The Huang-Rhys factors were calculated to confirm the existence of the high-frequency features observed in the spectral distributions of BoPTH. This was done by projecting the displacement in equilibrium geometries between the two states, $\mathbf{Q}^{(0,A)}-\mathbf{Q}^{(0,B)}$, into the normal mode basis of either molecule, and subsequently weighting by the normal mode frequencies, leading to the dimensionless weights:
\begin{equation}
    S_{AB}(\omega_\mu) = \frac{m_\mu\omega_\mu}{\hbar} \left(Q_\mu^{(0,A)}-Q_\mu^{(0,B)}\right)^2 
\end{equation}
Using the normal mode basis of either state yields very similar peak positions and intensities, as shown in Fig. \ref{fig:Huang-Rhys}, where it can also be verified that there is a collection of modes in the $\sim 3000 \ \wn$ region with large values of $S_{AB}$. 

\section*{Appendix E}
Profiles of the rates as a function of $\Delta F$ from Eq. \ref{nonrad_rate} are plotted in Figs. \ref{fig:BoANTH_rates} and \ref{fig:BoPTH_rates} for BoANTH and BoPTH respectively. Given that the figures are in log-linear axes, the classical (Marcus) rate behavior would be perfectly parabolic. It's apparent that the majority of these rate profiles differ substantially from being quadratic, indicating the importance of nuclear quantum effects, especially in the Marcus inverted regime. We also observe subtle but important differences between solvents in both molecules. 

\section*{References}
\bibliography{references}

\end{document}